\documentclass[pdflatex,sn-mathphys-num]{sn-jnl}%

\usepackage{graphicx}%
\usepackage{multirow}%
\usepackage{amsmath,amssymb,amsfonts}%
\usepackage{amsthm}%
\usepackage{mathrsfs}%
\usepackage[title]{appendix}%
\usepackage{xcolor}%
\usepackage{textcomp}%
\usepackage{manyfoot}%
\usepackage{booktabs}%
\usepackage{algorithm}%
\usepackage{algorithmicx}%
\usepackage{algpseudocode}%
\usepackage{listings}%
\usepackage{siunitx}%
\usepackage{gensymb}%
\usepackage{cleveref}
\usepackage{pdfpages}

\crefname{figure}{Fig.}{Figs.}
\crefname{equation}{Eq.}{Eqs.}
\crefname{proposition}{Proposition}{Propositions}

\let\origcite\cite
\def\cite#1{\unskip~\origcite{#1}}
\let\origcitep\citep
\def\citep#1{\unskip~\origcitep{#1}}

\theoremstyle{thmstyleone}%

\theoremstyle{thmstyletwo}%

\theoremstyle{thmstylethree}%

\let\vec\mathbf

\begin{document}

\title[Article Title]{Unidirectional information flow in a nanomagnetic metamaterial}

\author*[1]{\fnm{Johannes} \sur{H. Jensen}}
\email{johannes.jensen@ntnu.no}
\equalcont{These authors contributed equally to this work.}

\author[1]{\fnm{Ida} \sur{Breivik}}
\equalcont{These authors contributed equally to this work.}

\author[1]{\fnm{Arthur} \sur{Penty}}
\author[1]{\fnm{Anders} \sur{Str\o mberg}}
\author[1]{\fnm{Henrik} \sur{Tidemann Kaarbø}}
\author[1]{\fnm{Dheerendra} \sur{S. Bhandari}}
\author[1]{\fnm{Thea} \sur{M. Dale}}
\author[2]{\fnm{Michael} \sur{Foerster}}
\author[2]{\fnm{Miguel Angel} \sur{Ni\~no}}
\author[2]{\fnm{Deepak} \sur{Dagur}}
\author[1]{\fnm{Magnus} \sur{Sj\"alander}}
\author[1]{\fnm{Gunnar} \sur{Tufte}}
\author[1]{\fnm{Erik} \sur{Folven}}

\affil[1]{\orgname{Norwegian University of Science and Technology}, \city{Trondheim}, \country{Norway}}

\affil[2]{\orgname{ALBA Synchrotron Light Facility}, \orgaddress{\street{Carrer de la Llum 2 -- 26}, \city{Cerdanyola del Vallés}, \postcode{08290}, \state{Barcelona}, \country{Spain}}}

\abstract{

Artificial spin ice (ASI) are metamaterials composed of interacting nanomagnets.
Although ASI hold promise for low-power computing, the ability to transmit information through these two-dimensional systems has been limited. 
Inspired by non-reciprocal transport in nature, we develop a framework for non-reciprocal influence between nanomagnets.
Using the framework we discover a family of ASI geometries with inherent directionality. 
Directional ASI have the property that, when driven by an external field protocol, domains grow and reverse in the same direction, illustrating an emergent non-reciprocity of the system.
Combining growth and reversal results in unidirectional domain movement through the metamaterial.
We focus on one member of the directional ASI family, and demonstrate unidirectional domain growth experimentally. 
Furthermore, we show that the direction of growth is reconfigurable by tuning the external field strengths.
Finally, we demonstrate how the directionality of the system significantly improves memory capabilities in a reservoir computing framework.
Our work is the first demonstration of an ASI with inherent directionality, offering a magnetic computing platform that combines memory and computation within a single neuromorphic substrate.

}


\maketitle
\section*{Introduction}\label{sec:introduction}

Many systems in nature have an inherent direction, enabling reliable transport of energy, matter or information. 
For example, ion pumps in biological membranes enable ions to cross the membrane one way but not the other.
Directionality is also essential in engineered systems, e.g., digital computing is rooted in the unidirectional conduction of the p-n junction. 
Non-reciprocity is closely related to directionality, and has been reported in a range of optical~\citep{floess_tunable_2015, wang_self-induced_2025}, mechanical~\citep{brandenbourger_non-reciprocal_2019, chen_realization_2021, veenstra_non-reciprocal_2024}, superconducting~\citep{pal_josephson_2022, nagata_field-free_2025} and magnonic~\citep{chen_excitation_2019, jiang_realizing_2025, xu_geometry-induced_2025} systems. 
All of these systems rely on broken inversion symmetries, giving the systems an inherent sense of ``forward'' and ``backward''. 

Artificial spin ice (ASI) are metamaterials composed of interacting nanomagnets arranged on a two-dimensional lattice, which exhibit a wide variety of emergent behavior~\citep{Skjervo2020}.
Non-reciprocity, however, has not yet been explored in ASI systems.
Although ASI have been proposed as viable substrates for low-power neuromorphic computing~\citep{jensen_reservoir_2020, gartside_reconfigurable_2022}, the ability to transmit information within these two-dimensional systems has remained limited. 
Non-reciprocal phenomena in ASI could provide means for directional information transport, which would significantly increase their potential. 
Although unidirectional information transmission has been demonstrated in 1D chains of nanomagnets~\citep{nomura_controlling_2017}, it has not been extended to 2D systems.
More recently, we demonstrated how magnetic domains of specific shapes can break the symmetry in ASI and result in unidirectional domain propagation ~\citep{penty_controllable_2025}. 
Here, we consider symmetry breaking in the ASI geometry itself, resulting magnetic metamaterials with an \emph{intrinsic} directionality.

We present a family of directional ASI geometries, in which magnetic domains propagate in a single direction.
First, we develop a framework for non-reciprocal influences between nanomagnets, which we exploit to create ASI geometries where the total influences are directional.
We show that, when driven by an external field protocol, such geometries exhibit \emph{unidirectional} magnetic domain growth and reversal.
When growth and reversal is combined, magnetic domains propagate in a single direction through the metamaterial. 
Furthermore, tuning the field protocol can alter the direction of growth, suggesting a viable path towards reconfigurability. 
Finally, we demonstrate how directional ASI can be exploited for reservoir computing, where unidirectional domain movement enables significant memory capacity and computation in a single magnetic substrate.

\section*{Spin influence}

The dipolar interaction between two magnets is symmetric: the magnets exert an equal and opposite force on each other.
However, how the magnets react to a force can be asymmetric. 
When subject to a magnetic field, a nanomagnet will switch its magnetization if the field exceeds its switching threshold.
The switching threshold depends on the angle of the field, and can be described by a characteristic switching curve~\citep{stoner_mechanism_1948, breivik_tuning_2026}.
This angle-dependence can give rise to a non-reciprocal switching response, where one magnet influences another more than vice versa.

\newcommand{\hext}{\ensuremath{\vec{h}_{\text{ext}}}}
\newcommand{\hdip}{\ensuremath{\vec{h}_{\text{dip}}}}
\newcommand{\hdipij}{\ensuremath{\hdip^{ij}}}
\newcommand{\hi}{\ensuremath{\vec{h}_i}}

\begin{figure*}
\centering
\includegraphics[width=\textwidth]{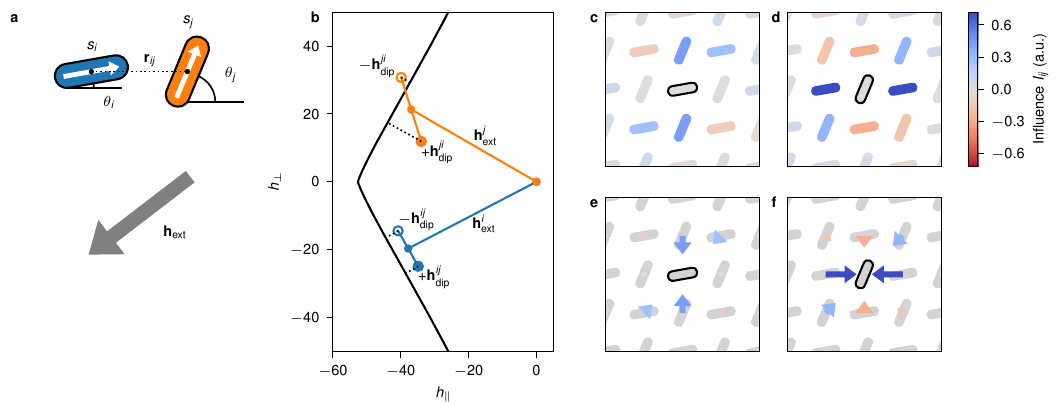}
\caption{\label{fig:influence}
\textbf{Spin influence.}
\textbf{a} Two nanomagnet spins $i$ (blue) and $j$ (orange), with states $s_i,s_j\in\{-1,1\}$, orientations $\theta_i$ and $\theta_j$, separated by some displacement $\vec{r}_{ij}$, subject to an external field \hext.
\textbf{b} Closeup of the switching astroid (black curve), and magnetic fields for spin $i$ (blue) and $j$ (orange).
Fields are decomposed into the parallel $h_\parallel$ and perpendicular $h_\perp$ components with respect to each spin axis, representing the local reference frame of each spin.
A spin will switch if subject to a magnetic field that crosses the astroid boundary.
The magnetic fields are composed of a global external field $\hext$ and the dipolar field from the other spin $\hdipij$.
Two possible dipolar fields are possible, depending on the state of the other spin, as indicated as closed circle ($s_j=+1$) and open circle ($s_j=-1$).
The dashed lines from the circles indicate the shortest distance to the astroid.
Note that the dipolar fields have been amplified for illustration purposes.
\textbf{c-d} Influence maps for a 2D system of spins based on \textbf{a}.
Neighbor spins are colored according to their influence on the center spin, where blue represents a stabilizing influence and red a destabilizing influence.
The influence has been normalized.
\textbf{e-f} Influence vectors for the spins in \textbf{c-d}.
The length of each vector corresponds to the influence from each neighbor.
}
\end{figure*}

To capture such asymmetries, we introduce \emph{spin influence}, which measures how much a magnet promotes or prevents the switching of another.
Here, we consider elongated single-domain nanomagnets which have two stable states (artificial spins).
In the following, we will refer to these nanomagnets simply as spins.
\cref{fig:influence}a shows two spins $i$ and $j$, with states $s_i,s_j\in\{-1,1\}$, orientations $\theta_i$ and $\theta_j$, separated by some displacement $\vec{r}_{ij}$.
In this example, a global external field $\hext$ is applied in the opposite direction of the spin states.

\Cref{fig:influence}b shows a closeup of the characteristic switching astroid of the nanomagnet spins~\cite{breivik_tuning_2026}.
A spin $i$ will flip (switch) if subject to a magnetic field $\vec{h}_i$ that crosses the astroid boundary (black curve).
In the figure, the fields are decomposed into $h_\parallel$ and $h_\perp$, which represent the field components parallel and perpendicular to the axis of each spin, respectively.
Thus, we are showing the astroid in the local reference frame of each spin.

In the isolated two-spin case, the total field $\vec{h}_i = \hext + \hdipij$, where $\hext$ is a global external magnetic field and $\hdipij$ is the dipolar field from spin $j$.
As illustrated by $\hext^i$ (blue) and $\hext^j$ (orange) in \cref{fig:influence}b, the external field is different in the local reference frames of each spin.
In other words, the spins experience a global \hext differently due to their relative orientation.

The direction of the dipolar field $\hdipij$ depends on the state of the neighbor spin $s_j=\pm1$.
Flipping spin $s_j\rightarrow-s_j$ results in a $180\degree$ rotation of the dipolar field ($\hdipij \rightarrow -\hdipij$).
These two possible dipolar fields are illustrated in \cref{fig:influence}a as closed circle ($s_j=+1$) and open circle ($s_j=-1$).
In the particular example shown in \cref{fig:influence}, we can see that $s_i=+1$ effectively makes spin $j$ (orange) harder to switch, by increasing the distance to the astroid boundary (dashed lines in the figure).
Conversely, if $s_i=-1$, the opposite is true, and the dipolar field effectively reduces the astroid distance, and would in this case push spin $j$ beyond the boundary and cause switching.
This is not true for the other spin $i$, however, where the dipolar field from $j$ neither increases nor reduces the distance to the astroid.
In this case, the dipolar field is parallel to the astroid boundary, meaning $s_i$ is effectively unaffected by the state of the neighbor spin $s_j$.
We have uncovered an \emph{asymmetry} between the two spins, where spin $i$ exerts a large influence on the state of spin $j$, whereas spin $j$ has no influence on the state of spin $i$.

Given this intuition, we define spin influence based on how the distance to the astroid changes, when considering the possible dipolar fields from a neighbor spin.
Let $d_i(\hi$) be the shortest distance from the total field $\hi$ to the astroid boundary for spin $i$ (dotted lines in \cref{fig:influence}b).
Furthermore, let $d_i$ be negative if $\hi$ is inside the astroid boundary, and positive outside.
Then, the spin influence $I_{ij}$ from spin $j$ to spin $i$ is

\begin{equation}
    I_{ij} = d_i(\hext - \hdipij) - d_i(\hext + \hdipij),
\end{equation}

where the first term corresponds to the neighbor state $s_j=-1$, and the second term corresponds to $s_j=+1$.

The magnitude of $I_{ij}$ describes how much a neighbor spin $j$ can affect the astroid distance for spin $i$.
By taking into account both possible neighbor states $s_j=\pm1$, the magnitude is made independent of the neighbor state $s_j$.
A positive $I_{ij}$ indicates that the dipolar field from $j$ nudges spin $i$ to take the same value, while a negative $I_{ij}$ indicates the opposite.
Non-reciprocity occurs when the influence between magnets is asymmetrical, i.e., $|I_{ji}|\ne|I_{ij}|$.
Asymmetries exist for many combinations of spin orientations (see Supplementary information).

Note that $I_{ij}$ is very much dependent on the state of the spin being influenced, since flipping $s_i$ will align the spin with $\hext$, making $h_\parallel$ positive, which is a stable configuration.
Furthermore, the astroid distance $d_i$ is only really meaningful when $h_i$ is close to the astroid boundary.

\cref{fig:influence}c-d depicts influence maps for a 2D geometry based on the non-reciprocal pair of spins in \cref{fig:influence}a.
As can be seen, there is an asymmetry between the neighbor influence for the two sublattices; the spins with orientation $\theta_i$ (\cref{fig:influence}c) experience a weaker influence from their neighbors compared to spins with orientation $\theta_j$ (\cref{fig:influence}d).
However, the local non-reciprocity between pairs of spins cancels out on a larger scale due to the symmetries of the 2D geometry.
The system as a whole is thus still reciprocal.

\subsection*{Emergent ferromagnetism}

ASI geometries may exhibit collective ferromagnetism\citep{macedo_apparent_2018} or antiferromagnetism\citep{zhang_crystallites_2013}.
Experimental results indicate that ferromagnetic systems may be more robust compared to antiferromagnetic systems, which tend to more easily get stuck in local energy minima~\citep{jensen_clocked_2024,Bhandari2025}.

When the spins are aligned, we can interpret a positive influence as describing a ferromagnetic relationship between the spins, and a negative influence as an anti-ferromagnetic relationship.
In other words, a ferromagnetic relationship is when the state $s_j=s_i$ results in a dipolar field that acts to prevent switching.
Conversely, an antiferromagnetic relationship is when the state $s_j=s_i$ results in a dipolar field that promotes switching.

Influence can hence be used to measure the amount of ferromagnetic or antiferromagnetic interactions in a system of spins.
The per-spin influence $I_{i} = \sum_{j\ne i} I_{ij}$ provides an overall measure of the influences on a spin $i$.
In an ensemble of spins, the average per-spin influence $F$ provides a total measure of (anti)ferromagnetism:

\begin{equation}
    F = \frac{1}{N}\sum_i^N I_i
\end{equation}

For instance, the predominance of positive influence values in \cref{fig:influence}c-d indicates a ferromagnetic bias and hence a positive $F$.

\subsection*{Directionality}

We have seen how spin influence provides a measure of how much one spin affects the state of another.
Here, we extend the measure to capture the direction of influence.

Given some spin $i$ and neighbor $j$, the influence vector from $j\rightarrow i$ is given by

\begin{equation}
    \vec{I}_{ij} = |I_{ij}| \vec{\hat{r}}_{ij},
\end{equation}

where $\vec{\hat{r}}_{ij}$ is the normalized distance vector from spin $j$ to $i$.
In other words, $\vec{I}_{ij}$ is simply the distance vector rescaled by the influence $I_{ij}$.

The per-spin influence vector $\vec{I}_i=\sum_{j\ne i} \vec{I}_{ij}$ indicates the direction from which a spin $i$ is most influenced.
The magnitude of $\vec{I}_i$ is a measure of directional asymmetry in the influences on spin $i$.
A zero magnitude indicates that the influence is symmetric, e.g., if spin $i$ is influenced equally by neighbors from all directions.
A positive magnitude, on the other hand, indicates some amount of spatial asymmetry is present, where spin $i$ is influenced more from one direction than the others.

Finally, the directionality vector $\vec{D}$ provides an overall measure of directionality in an ensemble of spins:

\begin{equation}
    \vec{D} = \sum_{i} \vec{I}_{i}.
\end{equation}

This provides a metric to evaluate and compare different ASI geometries.

\Cref{fig:influence}e-f depicts the influence vectors corresponding to \cref{fig:influence}c-d. 
Due to the symmetries of the geometry, all influence vectors will eventually cancel, resulting in a total directionality vector $\vec{D}=\vec{0}$.
A directional geometry is characterized by a nonzero $\vec{D}$, which we explore next.

\section*{Directional ASI geometries}

Based on spin influence, we investigate whether it is possible to have a directional ASI geometry.
In other words, is there an arrangement of nanomagnets that tile the plane and results in a directional asymmetry in the spin influences?

\begin{figure*}
\centering
\includegraphics[width=\textwidth]{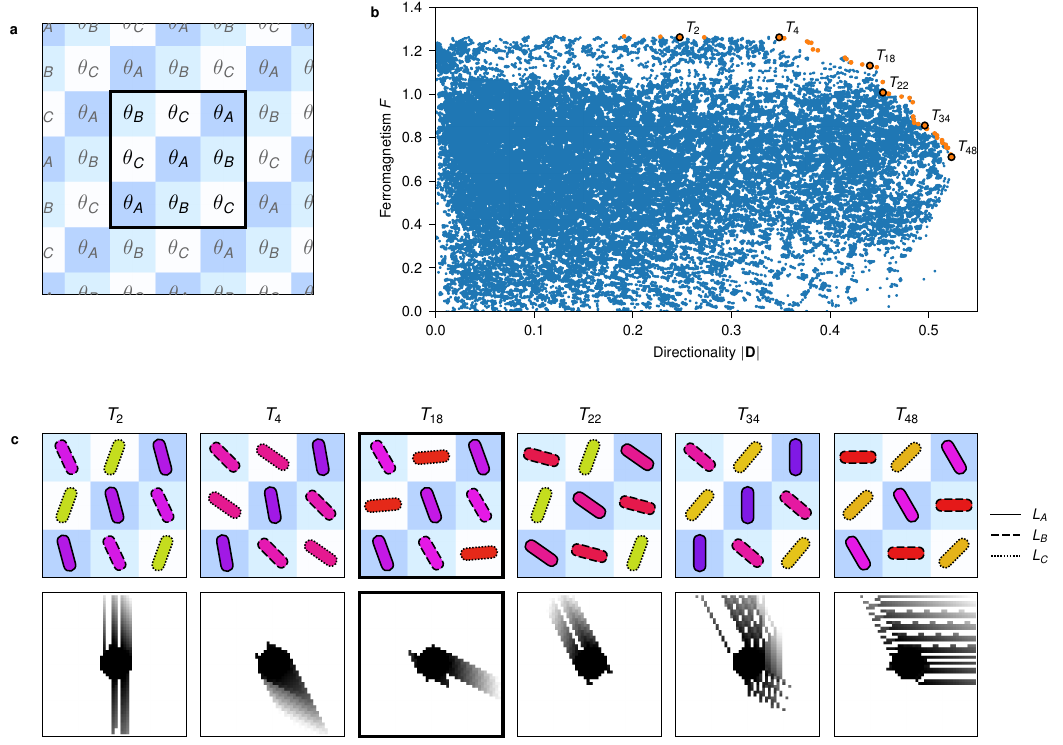}
\caption{\label{fig:tile}
\textbf{Directional ASI family.}
\textbf{a} Directional tile template.
The tile template is composed of three magnet orientations $\{\theta_A, \theta_B, \theta_C\}$, arranged along the diagonal.
\textbf{b} Space of directional geometries.
All combinations of magnet orientations are considered, plotted according to their directionality $|\vec{D}|$ and ferromagnetism $F$.
Each point in the plot corresponds to a $12 \times 12$ geometry and a three clock fields.
The number of possible clock fields per tile depends on the magnet orientations (see Methods).
The Pareto front of the search space is highlighted in orange, with six example tiles highlighted as $T_i$ according to the order they appear along the Pareto front.
\textbf{c} Tiles and their corresponding time-evolution.
Six example tiles from the Pareto front are shown.
The top row depicts the tile, while the bottom row shows time-evolution maps of the corresponding geometry.
An initial center domain is gradually grown over time using astroid clocking.
The shade of gray indicates the time step of growth, from black at initialization time to light gray towards the end, with white indicating no activity.
}
\end{figure*}

In this work, we consider a tile-based geometry on the square lattice, as depicted in \cref{fig:tile}a.
The geometry is constructed by tiling the square lattice with a $W\times H$ tile template.
The tile template defines the spin orientations $\theta_i$ of the geometry.
We allow the tile to contain at most three unique orientations (three sublattices).
With these constraints, the geometry is defined by a set of orientations $\{\theta_k\}$ and a mapping of the orientations onto the tile template.

Influence depends on the applied external field, in addition to spin orientations.
We consider astroid clocking to drive domain dynamics, where a set of field pulses selectively switch magnets from each sublattice~\citep{jensen_clocked_2024}.
Each sublattice in the system is therefore associated with a specific external clock field with angle $\phi_i$ and strength $H_i$.

Using stochastic optimization, we first search for a tile template that maximizes directionality $|\vec{D}|$ (see Methods).
The resulting tile is shown in \cref{fig:tile}a, which we refer to as the directional tile template, with the three spin orientations $\{\theta_A, \theta_B, \theta_C\}$ arranged along the diagonal in an alternating fashion.
Repeated optimization always results in some variant of the directional tile template.
Furthermore, optimization fails ($|\vec{D}| \approx 0$) with fewer than three angles or a smaller tile.
In the Supplementary information, we outline a proof of why the directional tile template maximizes directionality.
Notice how the resulting geometry is rotational asymmetric.

We first explore the possible geometries that arise from the directional tile template (\cref{fig:tile}a).
Specifically, we are interested in the interplay between directionality ($|\vec{D}|$) and ferromagnetism ($F$).
\cref{fig:tile}b shows the result of exhaustively searching all angle combinations $\{\theta_A, \theta_B, \theta_C\}$ (see Methods).
Each point in the plot corresponds to a geometry and a combination of clock fields (the number of clock field combinations per tile depends on the set of spin orientations $\{\theta_i\}$, see Methods).
As can be seen, there is an apparent tradeoff between directionality ($|\vec{D}|$) and ferromagnetism ($F$).
As directionality increases beyond $0.35$, there is a marked drop in ferromagnetism.
At maximum directionality, ferromagnetism is roughly half of its maximum.

The Pareto front of \cref{fig:tile}b is highlighted in orange, from which we have sampled some tiles which are shown in \cref{fig:tile}c.
Tiles are labeled as $T_i$ according to the order they appear along the Pareto front.
The top row of \cref{fig:tile}c shows the tile, whereas the bottom row depicts the domain evolution of the corresponding geometry.
Each geometry starts with an initial center domain, which is gradually grown using astroid clocking.
The shade of black indicates the time step of growth, from black at initialization time to light gray at the end, with white indicating no activity.
A variety of different behaviors are observed, ranging from symmetric ferromagnetic growth ($T_2$), directional ferromagnetic growth ($T_4,T_{18},T_{22}$) to directional anti-ferromagnetic growth ($T_{34},T_{48}$).

Most of the geometries along the Pareto front display some degree of directional domain growth.
The most directional geometries ($T_{34},T_{48}$) exhibit anti-ferromagnetic order, and tend to get stuck in local energy minima, prohibiting further domain growth.
In contrast, the directional ferromagnetic geometries ($T_{4},T_{18}$ and $T_{22}$) appear to offer more robust domain growth, while at the same time displaying directional behavior.
Our measures of directionality and ferromagnetism correspond fairly well to dynamic behavior, where directional growth is more prevalent at higher $|\vec{D}|$ values, and anti-ferromagnetic domains are more prevalent at lower $F$ values.
However, we also observe behavior that contradicts our measures, such as anti-ferromagnetic domains in a seemingly ferromagnetic system ($T_2$).
Such behavior can be explained by details in the neighbor influences, which are not captured in the overall influence measures.
In any case, the directional tile template results in a family of ASI geometries with a rich variety of emergent behavior.

\section*{The $T_{18}$ directional ASI}

\begin{figure}[h]
\centering
\includegraphics[width=\textwidth]{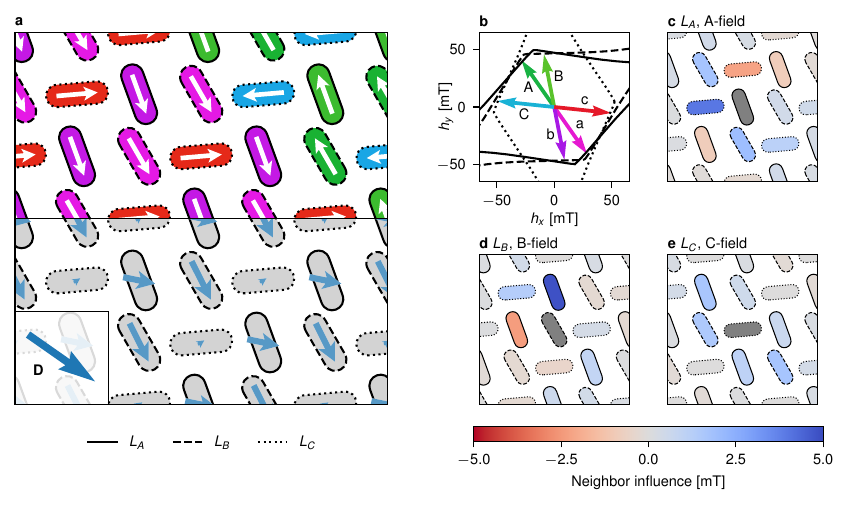}
\caption{\label{fig:i18}
\textbf{The $T_{18}$ geometry.} 
\textbf{a} A close-up of a $T_{18}$ ASI.
The three sublattices, $L_A$, $L_B$ and $L_C$ are marked with solid, dashed and dotted outlines, respectively. 
In the top half, the colors correspond to the state of the magnets, as indicated by the white arrows. 
In the bottom half, the per spin influence vector $\vec{I}_i$ is shown for each magnet. 
The total directionality vector $\vec{D}$ is shown in the inset. 
\textbf{b} The six clock fields used to grow ($A$, $B$ and $C$) and shrink ($a$, $b$ and $c$) domains. 
The switching astroids of the three sublattices are also shown. 
\textbf{c-e} Influence maps for the reference magnet (gray) belonging to each of the three sublattices, subject to their respective clock field. 
The blue magnets have a stabilizing effect on the reference, while red magnets have a destabilizing effect.
The influence maps correspond to the polarized state. 
}
\end{figure}

For the remainder of this paper, we focus on the ASI geometry based on the $T_{18}$ tile shown in \cref{fig:tile}c, which appeared particularly promising in simulations.
The $T_{18}$ geometry is shown in \cref{fig:i18}, where the three sublattices $L_A$, $L_B$ and $L_C$ are highlighted by the solid, dashed and dotted outlines, respectively.
The corresponding angles $\theta_A = \ang{-70}$, $\theta_B = \ang{-60}$ and $\theta_C = \ang{5}$.
We refer to the state of the magnets by their color, as indicated in \cref{fig:i18}a(top). 
The system is polarized when all magnets are in their pink-red state.

\cref{fig:i18}a(bottom) shows the per spin influence vectors $\textbf{I}_i$ for the magnets. 
As can be seen, the magnets in the $L_A$ and $L_B$ sublattices experience a significantly directional influence, with magnitudes of around $\qty{3.5}{\milli\tesla}$ for $L_A$ and $\qty{4.6}{\milli\tesla}$ for $L_B$.
Magnets in sublattice $L_C$ also have a nonzero per spin influence, although it is considerably smaller at around $\qty{0.8}{\milli\tesla}$.
The net directionality $\vec{D}$ points in the southeasterly direction, as shown in the inset. 

To evaluate the $T_{18}$ geometry, we employ astroid clocking to grow and reverse domains in the array.
From the exhaustive sweep we obtain six clock fields, as shown in \cref{fig:i18}b.
We use the $A$, $B$ and $C$ fields for domain growth, and the $a$, $b$ and $c$ fields, of the same angles but with opposite polarity, for domain reversal. 
The $A$ and $a$ fields selectively switch magnets in the $L_A$ sublattice, and similarly for $B, b, C$ and $c$.

\cref{fig:i18}c-e shows influence maps for the sublattices $L_A$, $L_B$ and $L_C$ when subject to their respective clock fields. 
Since spin influence is predominantly positive in all three sublattices, the neighborhood has a net stabilizing effect on the reference magnet. 
Magnets in $L_A$ are are most strongly influenced by their western nearest neighbor, while magnets in $L_B$ are primarily influenced by their northern nearest neighbor.
In contrast, the influence is more evenly distributed for $L_C$ magnets, resulting in a less directional influence vector.

\begin{figure*}
\centering
\includegraphics[width=\textwidth]{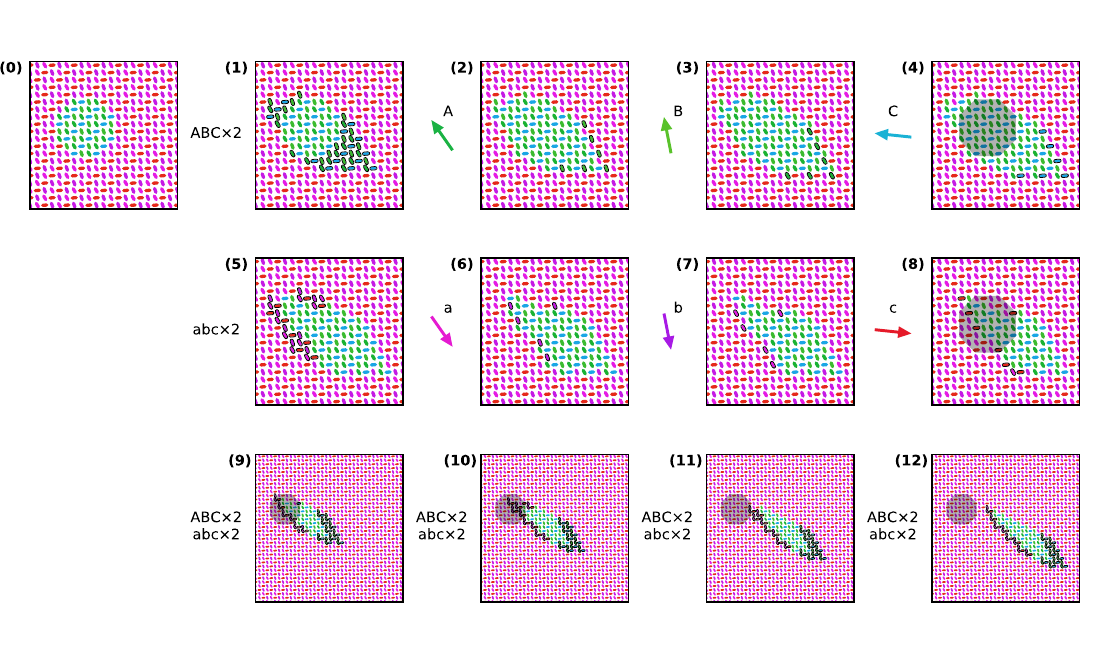}
\caption{\label{fig:SE-sim}
\textbf{Directional domain growth, reversal and movement.}
Starting from an initial domain (0), we apply three full $ABC$ clock cycles (1-4), growing the domain in the southeasterly direction. 
(1) shows the domain after the first two full $ABC$ cycles, while (2-4) show the domain after three subsequent field pulses, $A$, $B$ and $C$. 
We then apply three full $abc$ cycles (5-8), reversing the domain from the northwest. 
By alternating cycles of growth and reversal (9-12), we can move the domain in the southeast direction. 
The shaded regions highlight the initial domain.
}
\end{figure*}

\cref{fig:SE-sim} shows the time-evolution of the $T_{18}$ geometry during astroid clocking.
Snapshot 0 shows the initial state, an octagonal domain (blue-green) on a polarized background (pink-red). 
Snapshots 1-4 show how growth of the blue-green domain is predominantly in the southeasterly direction.
This directional growth is in agreement with $\vec{D}$.
Following growth, we reverse the domain using the $abc$ protocol, as shown in snapshots 5-8.
When reversing, notice how magnets along the northwestern edge of the domain switch.
As a result, the domain center shifts towards the southeast. 
Snapshots 9-12 demonstrate how alternating growth and reversal results in domain movement in the southeasterly direction. 
Note how the domain in snapshot 12 has clearly moved to the southeast from its initial state, as highlighted by the shaded region. 

The underlying mechanism behind the directional domain movement can be understood by considering the influence maps in \cref{fig:i18}c-e. 
We choose the $A$-field strength so that it is just strong enough to flip an isolated $L_A$ magnet. 
However, the influence from the neighboring magnets stabilize the $L_A$ magnets when in the pink-red state (\cref{fig:i18}c).
Consequently, $L_A$ magnets will only flip in close proximity to a blue-green domain.  
A blue western neighbor is sufficient to destabilize the $L_A$ magnet, which will then flip when the $A$ field is applied.
In other words, $L_A$ magnets within a pink-red domain are locked, but may be unlocked by first flipping their western nearest neighbor.

The $B$-field is chosen so that it is \textit{not} strong enough to flip isolated $L_B$ magnets. 
For the $L_B$ magnets to flip, the neighborhood must contribute with a significant destabilizing influence. 
As can be seen in \cref{fig:i18}d, the northern nearest neighbor must be green to unlock the $L_B$ magnets.
Hence the $A$-field can unlock $L_B$ magnets, which may flip by the subsequent $B$-field.

We choose the $C$-field in a similar manner to the $A$-field, so that it is strong enough to flip isolated $L_C$ magnets. 
Once again, in a pink-red domain, the influence from the neighboring magnets stabilizes the $L_C$ magnets.
However, as seen in \cref{fig:i18}c, the influence in the neighborhood of the $L_C$ magnets is more evenly distributed compared to $L_A$ magnets. 
Consequently, there are several combinations of neighboring magnets which may unlock $L_C$ magnets. 
In our simulations (\cref{fig:SE-sim}(4)), the $L_C$ magnets are unlocked by combinations of their western, northwestern and northern neighbors.

Only magnets along a domain wall can flip, as a result of the locking/unlocking mechanism of the $ABC$ clock protocol.
Furthermore, growth is limited to the southeasterly domain walls due to the asymmetrical distribution of influence, in particular for $L_A$ and $L_B$ magnets.
Repeated $ABC$ cycles thus results in step-wise domain growth in the southeasterly direction.

Note that the three clock fields should be applied in the $ABC$ order, due to the locking/unlocking mechanism.
Changing the order of the clock fields will slow down the domain growth, as some clock fields will be unable to switch locked magnets. 

Pure southeasterly growth of the domain is observed from the third clock cycle onward. 
In the two first clock cycles, the initial domain grows slightly to the northwest in addition to the growth in the southeast direction, as shown in \cref{fig:SE-sim}(1). 
However, the domain quickly reaches an equilibrium (\cref{fig:SE-sim}(2-4)), where only magnets along the southeastern edge of the domain flip, resulting in consistent growth in the southeasterly direction.

Domain reversal follows a similar mechanism to that of growth.
Reversal can be understood as growth of the outer pink-red domain in the southeasterly direction.
Due to the rotational symmetry of the clock fields and switching astroids, the influence maps for reversal are identical to those for growth (\cref{fig:i18}c-e). 
By following the same analysis as for growth, we see that the three reversal fields, applied in $abc$ order, will flip magnets along the northwestern edge of the blue-green domain, as shown in \cref{fig:SE-sim}(5-8).
Hence, domain growth and reversal will both move the center of mass of a blue-green domain in the southeasterly direction. 

Alternating growth and reversal produces domain movement, as in \cref{fig:SE-sim}(9-12) based on the same principles. 
To balance the activity during growth and reversal, we note that the $abc$ fields are all slightly smaller in magnitude then the $ABC$ fields, see Methods. 

The domain movement is not invariant under time-reversal, as the domain will not retrace its path if we reverse all magnet states and clock fields. 
When reversing the initial state and the clock fields, the domain moves in the same, southeasterly direction as before (see Supplementary information). 
Technically, we must also reverse the order of the clock protocol to exemplify true time-reversal. 
However, due to the unlocking mechanism of the clock fields, the clock fields which appear out of order have no effect on the system, resulting in the domain simply moving more slowly in the same, southeasterly direction. 
As the domain continues to move in the southeasterly direction, it can not return to its starting point by time reversal. 
We thus observe an emergent non-reciprocity in the domain dynamics of the directional ASI.
Interestingly, this emergent non-reciprocity is different from many other non-reciprocal systems, where reversing the magnetic fields also reverses the direction of the non-reciprocity~\citep{xu_geometry-induced_2025, pal_josephson_2022, floess_tunable_2015}. 

\section*{Experimental domain growth}

\begin{figure*}
\centering
\includegraphics[width=\textwidth]{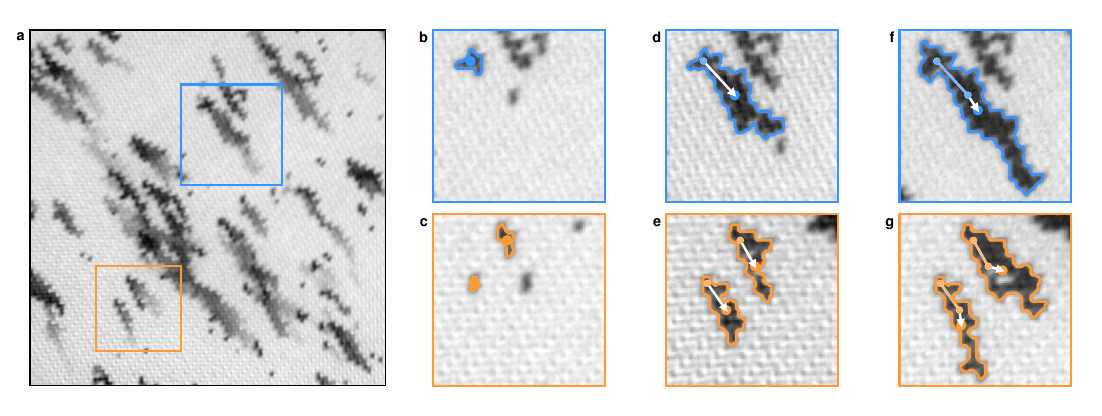}
\caption{\label{fig:SE-exp}
\textbf{Experimental directional domain growth of $T_{18}$.}
\textbf{a} Average intensity of 31 XMCD-PEEM images of 10 $ABC$ clock cycles (30 field pulses), starting from the polarized (white) state.
The intensity of the magnets indicate when the magnets switch in the series, where the dark magnets switch first and the light magnets switch last.
The imaged area is \qtyproduct{19x19}{\um}.
\textbf{b-g} Snapshots of the blue and orange regions of the array, at three intervals in the clock series: after \textbf{b-c} three, \textbf{d-e} six and \textbf{f-g} nine clock cycles.
The centers of mass of the tracked domains are marked by colored dots in all snapshots, and its movement from the previous snapshot is shown by the white arrows.
}
\end{figure*}

Next, we demonstrate directional growth experimentally.%
Between each applied clock field, we image the system using x-ray magnetic circular dichroism photoemission microscopy (XMCD-PEEM).
\cref{fig:SE-exp}a shows the average intensity of 31 images from ten cycles of the $ABC$ clock protocol (see also Supplementary information for the full series).
Nanomagnets that switch first appear darkest, while nanomagnets that switch in later steps appear lighter. 
Thus, by following the gradient in the magnetic domains from dark to light, we can see that the domains predominantly grow in the southeasterly direction. 

In the first experimental clock cycle, single magnets nucleate several places in the array. 
Single magnet nucleation can be attributed to natural variations in the coercive fields of the nanomagnets.
When we apply additional clock cycles, domains start growing from these nucleation sites.
While growth is predominantly in the southeasterly direction, there is also some growth in other directions during the first few clock cycles.
In later clock cycles, growth in the southeasterly direction is clearly most prevalent.

\cref{fig:SE-exp}b-g shows two regions of the array (blue and orange in \cref{fig:SE-exp}a) at three intervals throughout the clock series. 
We start from the highlighted domains shown in b and c, which is three clock cycles after polarization. 
d-e and f-g show the domains after three and six more clock cycles, respectively.
The center of mass of each domain moves in the southeasterly direction throughout the series, as indicated by the movement of the colored dots in b-g.

The three highlighted domains show some variation in their direction of growth throughout the series.
As there is some variation in the coercivity of the magnets, the experimental domains may stop growing at magnetically hard magnets, or grow by multiple magnets at magnetically soft magnets. 
Consequently, the experimental domains appear more organic than the ones observed in simulation.
We also observe deviations in the direction of growth when domains grow into each other. 
Nevertheless, the predominance of southeasterly growth demonstrates that directional growth in $T_{18}$ is robust to experimental conditions. 

When reversing the clock fields in experiment, we are not able to reproduce the clear reversal from the northwesterly direction as seen in simulation. 
The observed reversal regime is characterized by domains shrinking from all directions. 
This discrepancy between experiment and simulation suggests that there may be some aspects that are not captured by the dipole model.
For example, in tightly coupled nanomagnet arrays, dipole interactions may affect the uniformity of their magnetization, resulting in changes in their switching thresholds~\citep{leo_chiral_2021, strandqvist_nanomagnet_2025}. 
However, domain movement may still be possible using a combination of directional domain growth and symmetric domain reversal. 

Although the experimental southeasterly growth is only shown with a single set of field strengths, simulations indicate that similar results occur for a range of field strengths. 
The observed directional domain growth should thus be robust not only to natural variations of the coercive fields of the individual magnets, but also to variations in the experimental set-up, for example temperature effects. 
This further suggests that the $T_{18}$ geometry has potential in applications beyond the lab.

\section*{Northwesterly domain growth}

We now explore the possibility of driving the system outside the field range in which we observe southeasterly domain growth and movement. 
Increasing the field strengths could cause avalanches, where several nanomagnets will switch in a single clock field. 
While large avalanches will cause the growth to be come uncontrolled, small avalanches may be taken advantage of to create a wider range of possible domain dynamics. 

\begin{figure*}
\centering
\includegraphics[width=\textwidth]{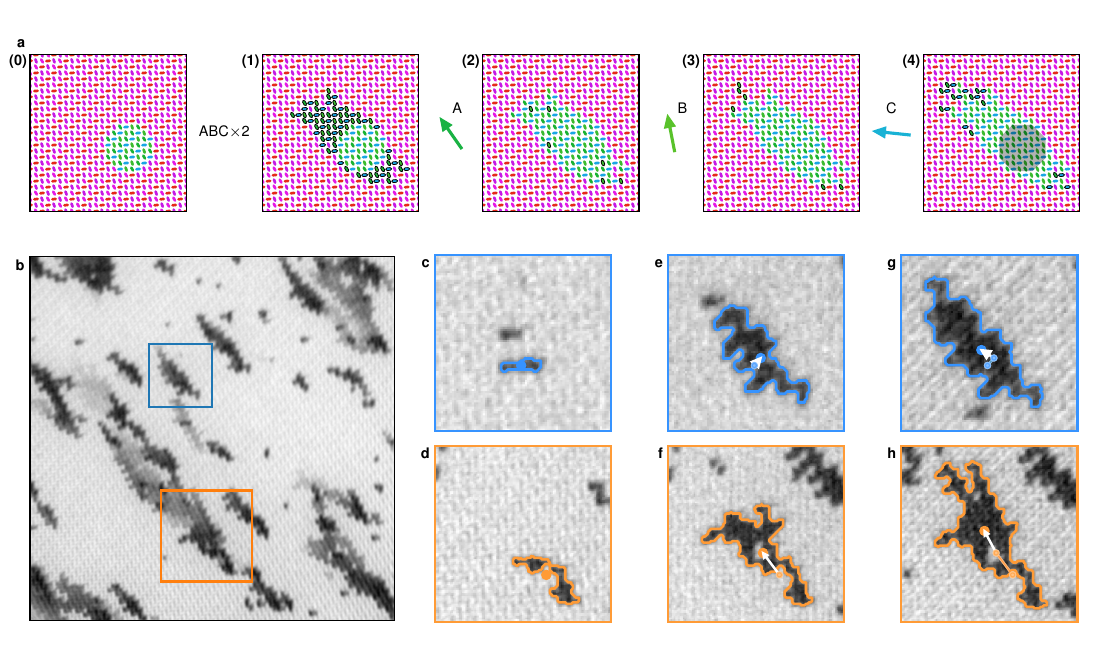}
\caption{\label{fig:NW}
\textbf{Northwesterly domain growth of $T_{18}$.}
\textbf{a} Simulated domain growth in the northwest regime. 
Starting from the initial state in (0), we apply three $ABC$ clock cycles (1-4).
The position of the initial domain is marked by the shaded region in (4).
\textbf{b-h} Experimental domain growth in the northwest regime.
\textbf{b} Average intensity of 31 XMCD-PEEM images of 10 full cycles (30 field pulses) of astroid clocking, starting from the polarized state.
The imaged area is \qtyproduct{19x19}{\um}.
\textbf{c-h} Snapshots of the blue and orange regions of the array, at three intervals in the series, after \textbf{c-d} one, \textbf{e-f} three and \textbf{g-h} five clock cycles.
The center of mass of the domain is marked by colored dots in all snapshots, and its movement from the previous snapshot is shown by the white arrows. }
\end{figure*}

Remarkably, we find a new growth regime in which the domains of $T_{18}$ grow in the northwesterly direction, i.e., in the \textit{opposite} direction as before. 
\cref{fig:NW}a shows a simulated series of domain growth in this regime. 
We find this growth regime when we apply weaker \textit{A}-fields and stronger \textit{B}- and \textit{C}-fields. 
In this regime, the \textit{B}- and \textit{C}-fields switch magnets in more than one sublattice, an effect we term ``overclocking''.

The first two clock cycles of the simulation (\cref{fig:NW}a(1)) are characterized by avalanches that switch a large area of nanomagnets to the northwest of the initial domain. 
This avalanche leaves behind some holes (unflipped $L_A$ nanomagnets) which are filled in by the subsequent \textit{A}-field in \cref{fig:NW}a(2). 
Next, the \textit{B}-field flips a few magnets along the northern edge of the domain, including a small avalanche consisting of magnets from both $L_A$ and $L_B$. 
When the \textit{C}-field is applied in \cref{fig:NW}(4), magnets in both the $L_B$ and $L_C$ sublattices flip. 
The resulting domain after another full clock cycle has grown mainly in the northwest direction, but still has some holes which can be filled by more clock cycles.

We are able to experimentally reproduce the northwest growth, as seen from the direction of the gradients in the average XMCD-PEEM image in \cref{fig:NW}b. 
For the full clock series, see Supplementary information. 
The experimental demonstration of northwesterly growth in the same array as the southeasterly growth rules out that systematic fabrication defects, e.g., causing a gradient in the coercivity of the nanomagnets in the array, contribute to the directional growth.

In \cref{fig:NW}c-h, we take a closer look at two regions of the array. 
Starting from the domains shown in c and d, one clock cycle after polarization, we apply two clock cycles, resulting in the domains shown in e-f. 
Applying two more clock cycles results in the domains shown in g-h. 
The center of mass of the domain marked in orange moves steadily northwest as the domain grows. 
The domain highlighted in blue also grows generally in the northwesterly direction, but not in a straight line. 
As growth in the northwest direction is dependent on small avalanches, it also seems more prone to uncontrolled switching in other directions, such as the growth to the east between c and e. 

When overclocking, growth in the northwest direction is faster than in the southeast direction, as the \textit{B} and \textit{C}-fields also flip magnets outside their corresponding sublattices.
This is apparent from the experimental data, where clocking the blue domain in the southeast regime for 9 clock cycles results in \cref{fig:SE-exp}f, which is approximately the same size as the orange domain after only 5 clock cycles in the northwest regime (\cref{fig:NW}h). 

In this case, the \textit{B} and \textit{C}-fields overclock the $T_{18}$ geometry, and the \textit{A}-field fills in the remaining nanomagnets. 
For this particular choice of fields, the resulting domain growth is in the northwest direction. 
However, by adjusting the field strengths, the direction of domain growth can be altered significantly. 

\section*{Phase space of directional growth}

By sweeping the field strengths of the three clock fields in simulation, we find a wide range of different types of domain growth in $T_{18}$, including directional and symmetric regimes.
We classify the growth regimes by tracking how the domain center of mass moves. 

\begin{figure*}
\centering
\includegraphics[width=\textwidth]{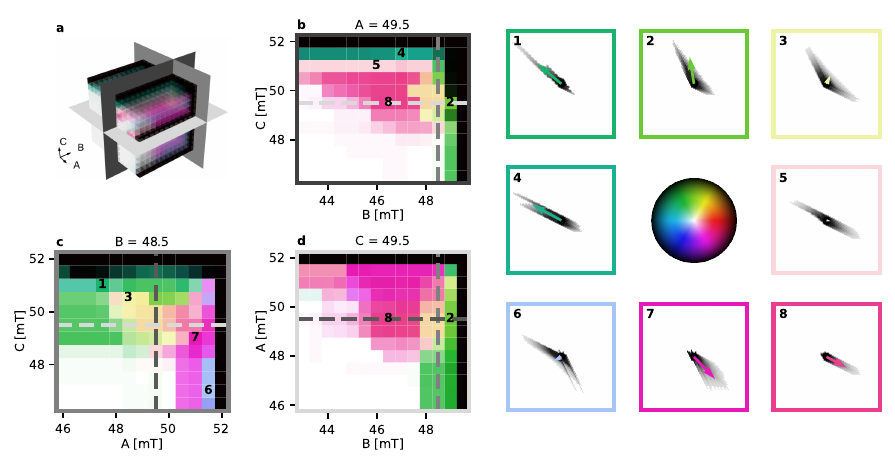}
\caption{\label{fig:phase_diagram}
\textbf{Phase space of directional growth.}
\textbf{a} Schematic of the 3D phase space of directional growth in the $T_{18}$ directional ASI, where hue represents direction and lightness represents the speed of growth (light is slow/no growth and dark is fast/avalanching growth).
The three slices shown in \textbf{b-d} are marked in gray in \textbf{a}.
\textbf{1-8} show examples of the variation of growth we find using varying field strengths, where the growth starts from the black initial domain and follows the gradient.
The colored arrow marks the direction of movement of the center of mass.
\textbf{1} and \textbf{8} are the northwest and southeast growth regimes, respectively.
}
\end{figure*}

\cref{fig:phase_diagram} shows three slices of the resulting phase diagram, where the color represents the direction of domain growth, and the lightness represents the growth speed. 
In the white areas of the phase diagram, the center of mass of the domain does not move, as the fields are too weak to flip any nanomagnets.
In the black areas, the fields are so strong that the domain quickly avalanches to the edge of the domain, growing in an uncontrolled manner. 
Between these two types of growth, we find a variety of growth speeds. 
Slow growth occurs when domains which stop growing after only a few clock cycles, or through symmetric growth, which will also result in little movement of the center of mass. 
Fast growth stems from overclocked regimes, where small avalanches make the domains grow faster than in regimes where only one sublattice switches at a time.

Remarkably, we can grow domains in almost any direction by adjusting the field strengths of the clock protocol. 
The right side of \cref{fig:phase_diagram} shows eight examples of domain growth found in the $T_{18}$ phase space. 
The northwest and southeast growth regimes are found in 1 and 8. 
We also find similar regimes in other regions of the phase space, such as the regimes shown in 2, 4 and 7. 
In 3 and 6, growth in the southeasterly and northwesterly direction is combined, resulting in movement of the center of mass in the northeasterly and southwesterly directions, respectively. 
There are also areas in the phase diagram where growth is less directional, such as in 5, where the domain grows only slightly more to the east than west, resulting in only a small change in the position of the center of mass. 
The wide range of growth regimes found in $T_18$ shows that the directionality of the system may be reconfigured simply by changing the field strength of the clock protocol. 

Unfortunately, the phase space for reversal is not as varied, and is largely dominated by reversal in the southeasterly direction (see Supplementary information).

\subsection*{Computation}

\begin{figure*}
\centering
\includegraphics[width=\textwidth]{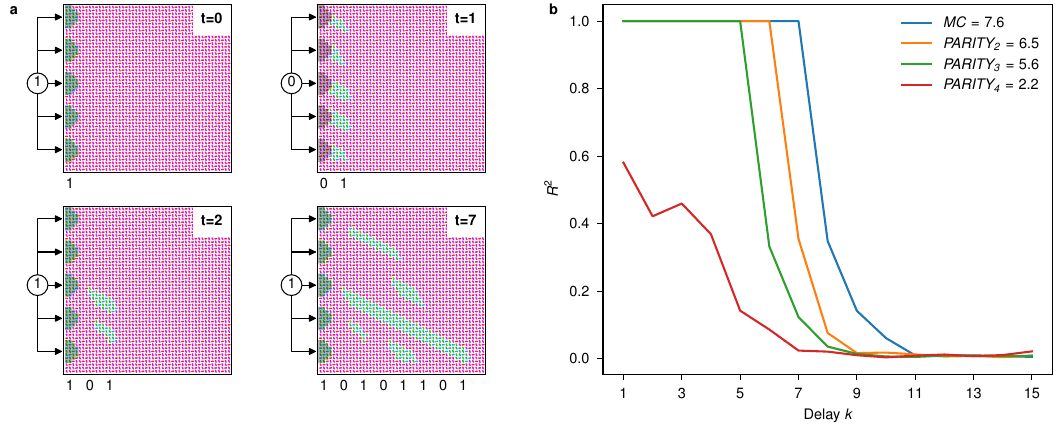}
\caption{\label{fig:computing}
\textbf{Reservoir computing with $T_{18}$.}
\textbf{a} Time evolution of $T_{18}$ subject to a binary input time series.
Input bits are encoded along the left edge, as indicated by the arrows into the shaded regions.
An astroid clocking protocol is applied between each input to move information southeast ($ABCabcabc\times3$).
Denoted below each snapshot are the input bits which have been encoded so far.
\textbf{b} Reservoir computing performance on two benchmark tasks: memory capacity ($\mathit{MC}$) and delayed $n$-bit parity ($\mathit{PARITY_n}$).
}
\end{figure*}

Finally, we demonstrate the utility of directionality for information processing, by considering $T_{18}$ in a reservoir computing framework~\citep{Jaeger2001,Maass2002}.
\cref{fig:computing}a shows simulation snapshots from our $T_{18}$ reservoir as it receives a binary input sequence.
As illustrated, each input bit is encoded as domains along the left edge of the $50\times50$ array.
A binary '1' is encoded as five circular blue-green domains on an pink-red background, whereas a '0' is encoded by reversing the circular regions to the pink-red state.
After input has been encoded ($t=0$), it is shifted (moved) one step southeast with astroid clocking ($t=1$).

We shift domains southeast with three cycles of the $ABCabcabc$ clock protocol.
The protocol has one step of growth ($ABC$) and two steps of reversal ($abcabc$) in order for the domains to maintain a consistent size.
With only one step of reversal (as in \cref{fig:SE-sim}), the domains would increase in size and over time overwrite any gaps representing '0' bits.

The bottom of each snapshot indicates the sequence of input bits that have been applied at the corresponding time step.
As the input domains propagate through the system, the reservoir maintains a fading memory of the input sequence ($t=2$).
With this particular reservoir size, input reaches the right edge of the array after seven steps ($t=7$).

We deliberately place the domains close together to facilitate interactions.
As we will see, domain interactions can be exploited for computations on the input sequence.
In some rows, domains only survive by merging with the next domain in the sequence, i.e., only when there are two subsequent '1' bits in the input.
For example, at $t=2$ the two topmost domains from $t=1$ have vanished as a consequence of the alternating input bits.
At $t=7$, however, an elongated domain survives in the second row from the top as the result of the bit sequence $11$.
In other words, domain interactions enable computations based on subsequent bits.
Note how each row of domains represent some transformation of the input sequence.

We evaluate the capabilities of the $T_{18}$ reservoir on a set of reservoir benchmark tasks, where the goal is to compute some function on the input bits after a variable delay $k$ (see Methods).
Each task thus requires a combination of memory and computation.
\cref{fig:computing}b shows task performance ($R^2$) as a function of delay $k$, where $R^2=1.0$ is a perfect score, and task performance is the sum of scores across delays.

The first task is memory capacity ($\mathit{MC}$), where the goal is to recover the input bit after each delay $k$, and is thus a measure of short-term memory.
As can be seen, $T_{18}$ has perfect memory for delays $k=1-7$.
$T_{18}$ maintains in its domains an input history of the previous seven input bits ($\mathit{MC}=7.6$).
Note that the $\mathit{MC}$ of the $T_{18}$ reservoir can be increased arbitrarily by scaling the width of the system, e.g., a $100\times100$ system obtains a $\mathit{MC}=15.4$ (see Supplementary information).
As a comparison, $\textrm{MC}\approx5$ was reported for pinwheel geometry\citep{Stenning2024}, whereas $\textrm{MC}\approx3.5$ was reported for a kagome system\citep{Hon_ASIrc2021}.

The second benchmark task is the delayed $n$-bit parity task, denoted by $\mathit{{PARITY_n}}$.
The goal of $\mathit{{PARITY_n}}$ is to compute the parity of $n$ subsequent bits after each delay $k$.
In contrast to $\mathit{MC}$, $\mathit{{PARITY_n}}$ requires a combination of short-term memory and computation (non-linearity) in the reservoir.
As seen in \cref{fig:computing}b, $T_{18}$ obtains excellent performance on both the $2$-bit and $3$-bit parity tasks.
Both tasks are bounded by the $\mathit{MC}$ of the system.
For $n=4$ bits and beyond, performance stagnates due to limited long range interactions between the input domains.

With the $T_{18}$ reservoir, we have demonstrated how unidirectional domain movement can be exploited for short-term memory and computation.
While we have previously demonstrated short-term memory in ASI with the ``snake'' glider~\citep{penty_controllable_2025}, the directional reservoir is the first to exhibit significant memory and computation capabilities within the same substrate.
The significant directionality of the $T_{18}$ geometry results in a reservoir in which memory is maximized.
However, maximizing memory comes at a cost of reduced computing capacity, due to the universal memory-nonlinearity tradeoff in reservoir computing~\citep{Dambre2012}.

Other dynamic regimes of $T_{18}$ may offer a different balance between memory and computing.
It is also likely that other directional geometries offer distinct capabilities, e.g., by considering tiles with lower directionality (\cref{fig:tile}).
Regardless, $T_{18}$ demonstrates the existence of ASI geometries with significant memory capacity.
The directionality continuum remains largely unexplored, and more work is needed to uncover the computational capabilities of directional ASI.

\section*{Discussion}\label{sec:discussion}

In this work, we have presented a family of ASI geometries with intrinsic directionality.
These systems have the property that, when driven by an external field protocol, domains will grow and reverse in a single direction.
Combining growth and reversal results in unidirectional domain movement.
In these systems, directionality emerges from non-reciprocal influence between the nanomagnets.
To the best of our knowledge, this is the first study of non-reciprocity in ASI.
It is also the first demonstration of domain translation in ASI without the need for specific domain shapes.

Compared to other non-reciprocal magnetic systems such as spin waves~\citep{xu_geometry-induced_2025}, directional ASI offer significant control over the time-scales of domain propagation.
Through astroid clocking, domain growth and reversal proceeds in discrete steps in response to each clock field, which means the process can be stopped and started at will.
Hence the speed of the growth and reversal process can be controlled directly by tuning the speed of the field protocol.
As a computing substrate, directional ASI offers information flow which can be readily adapted to different input time scales.

For the $T_{18}$ geometry, domain growth can be tuned by varying the field strengths of the external field protocol. 
Remarkably, we find growth regimes for a variety of alternate directions, even the opposite of the nominal southeasterly direction.
Field strengths have significant influence on the direction of growth, which presents a practical route towards reconfigurability.

Finally, we demonstrated how unidirectional domain movement can be exploited for information transmission and memory in a computing substrate.
Our $T_{18}$ reservoir computer obtains obtains a memory capacity which far exceeds previous ASI reservoirs.
Useful computations arise through domain interactions, offering a physical reservoir that combines memory and computing in a single substrate.
Information transmission is a fundamental computational property~\citep{Langton1990}, and our work readily applies to other computing paradigms such as nanomagnetic logic and neuromorphic computing.

The $T_{18}$ geometry explored here is only one member of a large family of directional ASI.
Within the family we find geometries with varying degrees of directionality.
Geometries with less directionality will offer a different balance between memory and nonlinearity, and might be better suited for less memory-intensive tasks.
In addition, tuning the field strengths or modifying the input encoding is sure to affect computational properties, presenting a viable route towards a reconfigurable magnetic computing substrate.

Beyond computing, directional ASI has a variety of potential applications, such as guiding magnetic nanoparticles along predefined paths\cite{gunnarsson_programmable_2005}, e.g., in lab-on-a-chip devices. 
The emergent non-reciprocity also suggests that directional ASI geometries may be viable as model systems for studying non-reciprocal phenomena.
Our results demonstrate the potential for unusual emergent properties when symmetry is systematically broken at the geometry level.

Ultimately, our work lays the foundation for information flow in ASI computing substrates.
As we have demonstrated, controlling information flow is crucial for the memory capabilities of such systems.
When memory is combined with the rich nonlinear dynamics of nanomagnetic metamaterials, the result is a highly potent all-magnetic substrate for ultra-low power computing devices.

\section*{Methods}\label{sec:methods}

\subsection*{flatspin simulations}
Simulations were carried out using the flatspin point-dipole ASI simulator~\citep{jensen_flatspin_2022}.

Switching parameters for the nanomagnets were $h_k=\SI{0.1303}{\tesla}$, $b=0.404$, $c=1.0$, $\beta=1.73$, and $\gamma=3.52$.
The values of $b, c, \beta, \gamma$ and $h_k$ were taken from the astroid database provided with flatspin, for stadium magnets of size \qtyproduct{220 x 80 x 10}{\nano\metre}, with a saturation magnetisation value of \SI{860}{\kilo\ampere\per\meter}. 
Other parameters include coupling strength $\alpha=0.0013$, a lattice spacing of $1$ and no disorder.%

For the initial evolutionary search and subsequent exhaustive search of the directional tile template (\cref{fig:tile}), systems of $12 \times 12$ nanomagnets were evaluated.
To avoid edge effects, a neighbor distance of $3$ was used in the evolutionary search and exhaustive sweep. 
For all other simulations, larger systems ($50\times 50$ for computing and $100\times100$ elsewhere) with a neighbor distance of $10$ were used. 
To prevent edge nucleation, we add three layers of buffer magnets (one $3\times3$ tile) around the edge of the array which are made harder to switch by doubling $h_k$.

In the simulations presented in \cref{fig:SE-sim} the directional ASI was clocked using clock fields with $\phi_A = \ang{-55}$, $\phi_B = \ang{-79}$ and $\phi_C = \ang{-7}$. 
For growth, we use $H_A = \qty{49.5}{\milli\tesla}$, $H_B = \qty{46.5}{\milli\tesla}$ and $H_C = \qty{49.5}{\milli\tesla}$. 
The domain is reversed using $H_a = \qty{49}{\milli\tesla}$, $H_b = \qty{46}{\milli\tesla}$ and $H_c = \qty{49}{\milli\tesla}$. 
To prevent switching from the edges, the three outermost layers of magnets (a full unit cell) have twice as large $h_k$. 
Domain growth in the northwesterly regime (\cref{fig:NW}) is done using a similar system and the same simulation parameters, but with $H_A = \qty{47.5}{\milli\tesla}$, $H_B = \qty{48.5}{\milli\tesla}$ and $H_C = \qty{51}{\milli\tesla}$. 

The phase diagram shown in \cref{fig:phase_diagram} is based sweeps of all three clock field strengths.%
The angles of the clock fields are the same for all phases. 

\subsection*{Influence}
When computing the influence $I_{ij}$ from the neighbors $j$ of each spin $i$, only spins within a radius of three lattice constants are considered neighbors ($28$ magnets).
The shortest distance to the astroid $d_i(\hi)$ is computed numerically.

\subsection*{Evolutionary search}
The $3 \times 3$ directional tile template (\cref{fig:tile}a) was discovered using a $(\mu+\lambda)$ evolutionary strategy (ES).
The ES begins with a random population of individuals (solutions).
For each generation of the algorithm, $\lambda$ offspring are produced from a random sample of the $\mu$ parents, and added to the population.
Offspring are produced as either a mutation of one parent, or a crossover of two parents.
Next, the fitness of the individuals in the population are evaluated (see below).
Finally, the best $\mu$ individuals are selected (according to their fitness) to survive to the next generation.
Thus, through a process akin to natural evolution, the population undergoes iterative improvement.
A population size of $100$ and $50$ generations was sufficient for convergence.
Other parameters include a mutation probability of $0.6$, crossover probability $0.4$, $\mu=50$ and $\lambda=50$.

Each individual represents a $W \times H$ tile, namely a set of $K$ spin orientations and a mapping from the set of orientations to each cell in the tile.
For the genome we use a simple list of $K + WH$ numbers to be optimized by the ES.
The set of spin orientations are constrained to the range $[-90\degree,90\degree]$.
To evaluate an individual, its tile is used as the basis of a $12\times12$ geometry.
Since there are $K$ spin orientations in the tile, the resulting geometry will have $K$ sublattices.
As fitness function we use the magnitude of the directionality vector $|\vec{D}|$ (higher is better).

For spin influence to capture (anti-)ferromagnetic interactions, spins should have their positive orientation ($s_i=+1$) aligned in the same general direction.
For example, a ferromagnetic interaction arises when two spins are positioned along their easy axis ($\vec{r}_{01}=[0, 1]$ with $\theta_1=\theta_2=90\degree$).
Since the meaning of $s_i=\pm1$ is arbitrary, we are free to rotate any spin by $180\degree$ without affecting the behavior of the system.
However, two spins with (near-)opposite orientations ($\theta_1=90\degree$ and $\theta_2=-90\degree$) would result in a negative influence $I_{ij}$ and incorrectly considered anti-ferromagnetic.
To counter such cases, we apply a simple algorithm to automatically detect a sensible ``spin axis'' for the geometry.
For the unique orientations $\theta_i$ in the geometry, we try all possible combinations of $180\degree$ rotations of the spins.
For each combination, we calculate the total magnitude of the resulting spin vectors.
Then we pick the combination with the highest magnitude, i.e., we rotate spins by $180\degree$ such that their positive orientation is aligned in the same general direction.

Influence depends both on the geometry and the external field $\hext$.
In astroid clocking, each sublattice is addressed by a specific clock field.
A tile is thus associated with $K$ clock fields, which are not known apriori.
The set of $K$ sublattices are associated with $K$ overlapping switching astroids (see \cref{fig:i18}b for an example with three astroids).
A sublattice $k$ is addressable iff there exists a gap between the astroids such that the astroid boundary for $k$ is inside all other astroids.
If this is the case, a clock field can be applied to exclusively switch spins from sublattice $k$ without affecting any other spins.
For each sublattice, we sample clock field candidates where the distance to the next (outer) astroid is at least \qty{2}{\milli\tesla}.
For each individual, we search over all possible combinations of clock field candidates, and use the combination which results in the highest fitness.

\subsection*{Exhaustive search for directional tiles}

We exhaustively search all possible spin orientations $\{\theta_A, \theta_B, \theta_C\}$ in the directional tile template (\cref{fig:tile}b).
Orientations are constrained to the range $[-90\degree,90\degree]$.
Due to translational and rotational symmetry of the tile, we only need to consider cases where $\theta_A < \theta_B < \theta_C$.

For each tile, we construct a $12 \times 12$ geometry and evaluate its ferromagnetism $F$ and directionality $|\vec{D}|$ (\cref{fig:tile}b).
We apply the same ``spin axis'' algorithm as described for the ES.
Similarly, for each geometry, we consider a range of potential clock fields and their combinations.
The number of clock fields depends on the spin orientations, i.e., the number of overlaps in the switching astroids.
Like the ES, we require clock fields to have at least \qty{2}{\milli\tesla} of spacing to the next (outer) astroid.
Hence, each point in \cref{fig:tile}b corresponds to a set of spin orientations $\{\theta_A, \theta_B, \theta_C\}$ and three clock field strenghts $\{H_A, H_B, H_C\}$.
The number of clock field combinations per tile range from $1-64$ with a mean of $21.5$.

The pareto front of \cref{fig:tile} is further investigated by simulating their time evolution with astroid clocking (bottom row of \cref{fig:tile}c).
Here we use a larger system of $50 \times 50$ spins, initialized with an octagonal domain in the center of an oppositely polarized background.
We employ a simple auto tuning algorithm which adjusts the strengths of the clock fields such that the number of spin flips per field is in the range $[2,12]$.
Then we clock the system with the tuned fields for $20$ clock cycles and record the time evolution.

\subsection*{Sample fabrication}
We fabricated a \numproduct{100x100} directional ASI with tile $T_{18}$. 
The ASI consists of \qtyproduct{220x80x10}{\nm} stadium-shaped nanomagnets with a lattice spacing of \qty{205}{\nm}. 
To prevent nucleation along the edges of the array, the three outermost layers of the ASI consist of \qtyproduct{220x70x10}{\nm} nanomagnets, which have a higher coercive field. 
The three outermost layers of nanomagnets are not included in \cref{fig:SE-exp}a or \cref{fig:NW}b.

Samples are fabricated using an electron beam lithography lift-off process. 
We coat the Si substrate with a \qty{95}{\nm} layer of electron resist, by spin-coating a 1:2 CSAR62:anisole mixture onto the substrate at \qty{2500}{rpm}.
The sample is soft-baked at \qty{150}{\degreeCelsius} for \qty{1}{\min}, before it is exposed using the Elionix ELS-G100 EBL system. 
We then develop the resist layer in AR600-546 e-beam developer for \qty{40}{\sec}. 
A \qty{10}{\nm} Permalloy ($\text{Ni}_{0.81}\text{Fe}_{0.19}$) layer, followed by a \qty{2}{\nm} Al layer, is deposited using the K.J. Lesker E-beam evaporator, and ultrasound-assisted lift-off is performed in AR600-71 resist remover. 

\subsection*{XMCD-PEEM and astroid clocking}
We perform astroid clocking experimentally by applying global field pulses, using an in-plane quadrupole magnet~\citep{foerster_custom_2016}. 
Between pulses, we image the magnetic state of the ASI using X-ray magnetic circular dichroism photoemission electron microscopy (XMCD-PEEM)~\citep{aballe_alba_2015}. 
We use the signal at the Fe L$_3$ edge for ferromagnetic XMCD contrast. 
To obtain sufficient magnetic contrast in all three sublattices of the ASI, we orient the array so that the incoming light is at an angle \ang{-41.6} with respect to the x-axis of the ASI, resulting in magnetic contrast along the same angle. 

We start from the polarized state, which we obtain by applying a \qty{-54.5}{\milli\tesla} field at a \ang{-42} angle.
We grow the magnetic domains by employing astroid clocking, where we apply a series of in-plane global field pulses along specific clock field angles.
For growth in the southeast direction, we use the field strengths $H_A = \qty{42.7}{\milli\tesla}$, $H_B = \qty{40}{\milli\tesla}$ and $H_C = \qty{42.7}{\milli\tesla}$. 
For growth in the northwest direction, we use the field strengths $H_A = \qty{40}{\milli\tesla}$, $H_B = \qty{40.9}{\milli\tesla}$ and $H_C = \qty{47.7}{\milli\tesla}$.

\subsection*{Computation}
For reservoir computing, we generate a random input sequence $u(t)$ of $1000$ bits as input.
Each input bit is applied at five circular regions along the left edge of the $T_{18}$ array (see \cref{fig:computing}a).
The five regions are separated by three rows of magnets, creating some diversity through different edge terminations.
A binary '1' is encoded by setting the five regions to the blue-green state, whereas a '0' is encoded by reversing the circular regions to the pink-red state.
The initial state of the array is fully polarized (pink-red state).

Between each input bit, the encoded domains are shifted southeast with three cycles of the $ABCabcabc$ clock protocol.
Field strengths were tuned to allow robust propagation of these smaller domains: 
$H_A = \qty{49.0}{\milli\tesla}$,
$H_B = \qty{47.4}{\milli\tesla}$,
$H_C = \qty{49.7}{\milli\tesla}$,
$H_a = \qty{50.3}{\milli\tesla}$,
$H_b = \qty{48.3}{\milli\tesla}$ and
$H_c = \qty{48.3}{\milli\tesla}$.

As reservoir output $\vec{x}(t)$, we sample the state of all the magnets after the three clock cycles, i.e., immediately before the next input bit is applied.

The trained output $y_\mathrm{train}=W_\mathrm{out} \vec{x}(t)$, where $W_\mathrm{out}$ is obtained with ridge regression using regularization parameter $\alpha=1.0$.
Performance is measured as the Pearson correlation coefficient ($R^2$) between the trained output $\hat{y}(t)$ and the target signal $y(t)$.
Ten-fold cross-validation was used on the entire reservoir output sequence $\vec{x}(t)$, and we report mean $R^2$ over the ten-fold CV.

Memory capacity $\mathrm{MC}$ measures the ability of the reservoir to recall delayed versions of the input sequence $u(t)$.
For a given delay $k$ the target signal is simply $y_k(t)=u(t-k)$, and the corresponding memory capacity $\mathrm{MC}_k$ is simply the $R^2$ for the target he delay $k$.
The total memory capacity is then $\mathrm{MC}=\sum_k\mathrm{MC}_k$.

The delayed $n$-bit parity $\mathrm{PARITY}_n$ is similar to memory capacity, but requires both memory and computation.
Here, the target signal is $y_k(t)=u(t-k-1) \oplus u(t-k-2) \oplus \cdots \oplus u(t-k-n)$, i.e., $y_k(t)=1$ if the last $n$ bits had an odd number of $1$'s and $y_k(t)=0$ otherwise.

\section*{Acknowledgments}
Experiments were performed at the CIRCE beamline at ALBA Synchrotron with the collaboration of ALBA staff. 
This work was funded in part by the Norwegian Research Council TEKNOKONVERGENS project SPrINTER (Grant no. 331821), and in part by the EU FET-Open RIA project SpinENGINE (Grant no. 861618).
Simulations were executed on the NTNU EPIC compute cluster\cite{Epic2019}.
The Research Council of Norway is acknowledged for the support to the Norwegian Micro- and Nano-Fabrication Facility, NorFab, project number 349807.

\section*{Author contributions}
JHJ and IB conceived and designed the study and contributed equally to this work.
JHJ developed the spin influence framework, discovered the directional geometries and did the theoretical study, in collaboration with AP.
IB fabricated the samples, assisted by TMD.
IB, AS, DSB and TMD conducted initial experiments.
IB led the XMCD-PEEM experiments and AP, HTK, DSB, TMD, GT and EF contributed to these measurements.
MF, MAN and DD provided support during XMCD-PEEM measurements and PEEM vector magnet operation.
IB performed the phase space exploration.
JHJ performed the reservoir computing study.
HTK provided the mathematical proof for the directional tile template. 
MS, GT and EF oversaw the project and provided feedback and direction throughout.
JHJ and IB wrote the manuscript with input from all authors.

\section*{Competing interests}
The authors declare no competing interests.

\bibliography{references}%



\section*{Supplementary material}

# Supplementary Discussion

## Directional tile maximizes directionality

The aim of this section is to illuminate why the directional tile maximizes directionality, and why directionality requires at least three magnet orientations.

**Proposition 1** *A directional ASI must have broken inversion symmetry.*

The spin influence from magnet $j$ onto magnet $i$ is denoted by $I_{ij}$, and given by

$$I_{ij} = d_i(\mathbf{h}_{\text{ext}} - \mathbf{h}_{\text{dip}}^{ij}) - d_i(\mathbf{h}_{\text{ext}} + \mathbf{h}_{\text{dip}}^{ij}), \quad (1)$$

where $d_i(\mathbf{h})$ is a scalar function fully determined by the properties of magnet $i$, $\mathbf{h}_{\text{ext}}$ is the external magnetic field, and $\mathbf{h}_{\text{dip}}^{ij}$ is the magnetic dipole field generated by magnet $j$ at the position of magnet $i$. Explicitly, the dipole field $\mathbf{h}_{\text{dip}}^{ij}$ is given by

$$\mathbf{h}_{\text{dip}}^{ij} = \alpha \sum_{j \neq i} \frac{3\mathbf{r}_{ij}(\mathbf{m}_j \cdot \mathbf{r}_{ij})}{|\mathbf{r}_{ij}|^5} - \frac{\mathbf{m}_j}{|\mathbf{r}_{ij}|^3}, \quad (2)$$

where $\mathbf{r}_{ij} = \mathbf{r}_i - \mathbf{r}_j$ is the distance vector from magnet $i \to j$, $\mathbf{m}_j$ is the unit vector parallel to the magnetization of magnet $j$, and $\alpha$ scales the dipolar coupling strength. Furthermore, the influence vector from $j \to i$ is given by

$$\mathbf{I}_{ij} = |I_{ij}|\hat{\mathbf{r}}_{ij}. \quad (3)$$

The inversion $r_{ij} \to -r_{ij}$ leaves Eq. (2) and consequently $I_{ij}$ unchanged. Therefore, $\mathbf{I}_{ij}$ is antisymmetric under the same inversion. Thus, the per-spin influence vector $\mathbf{I}_i = \sum \mathbf{I}_{ij} = 0$ for any spin in an infinite ASI with inversion symmetry. While edge effects may result in local directionality in a finite system, we disregard these small effects here. An ASI with net directionality $\mathbf{D} = \sum \mathbf{I}_i \neq \mathbf{0}$ must therefore have broken inversion symmetry.

**Proposition 2** *A directional ASI based on the square lattice must have more than two nanomagnet orientations.*

First, we consider a 1D chain of nanomagnets with only two nanomagnet orientations, $A$ and $B$. When traversing the 1D chain, being at a magnet of type $A$, there are only two possibilities for the next magnet in the chain, $A$ or $B$. The combination $AA$ is trivially mirror symmetric, causing directional cancellation, and thus, we are forced to have at least one $AB$ pairing for there to be nonzero directionality. However, as crystals are periodic, and we started at a magnetic type $A$, we know that we must at some point return to an $A$ type from a $B$ type. Thus, in a crystal, for each $A \to B$ pair, we must have a $B \to A$ pair. Due to the antisymmetry of $\mathbf{I}_{ij}$, any directional influence contribution made by the $AB$-pair will exactly cancel the contribution from

the $BA$-pair. Thus, the net directionality **D** in a 1D crystal with only two nanomagnet orientations *must* sum to zero.

Notably, the seemingly 1D nanomagnet shift register [1] does have a net directionality. However, the arrangement of nanomagnets break inversion symmetry by spatially offsetting one sublattice in the second dimension, so that $I_{AB} \neq I_{BA}$. It is thus not truly 1D.

The 2D square lattice can be considered as a cartesian product of 1D chains. Therefore, in a square lattice with two magnet orientations, there can be no accumulation of directionality. The argument above holds for any line drawn through the points of the lattice. By adding a third nanomagnet orientation to the system, it is possible to break the local symmetries preventing accumulation of directionality.

**Proposition 3** *Any $3 \times 3$ directional tile must follow the directional tile template.*

There are several ways of placing nanomagnets with three separate orientations on a $3 \times 3$ square tile. However, to maximize directionality, we should be able to accumulate directionality (in the same direction) in all rows and all columns of the tile. Thus, we must place the nanomagnets so that all rows and all columns of the tile include all three nanomagnet orientations (satisfying theorem 2), and so that the final tile has no inversion symmetry (satisfying theorem 1). Interestingly, the only way of achieving this is by placing the nanomagnets as specified by the directional tile template.

# Supplementary Figures

## Influence asymmetry

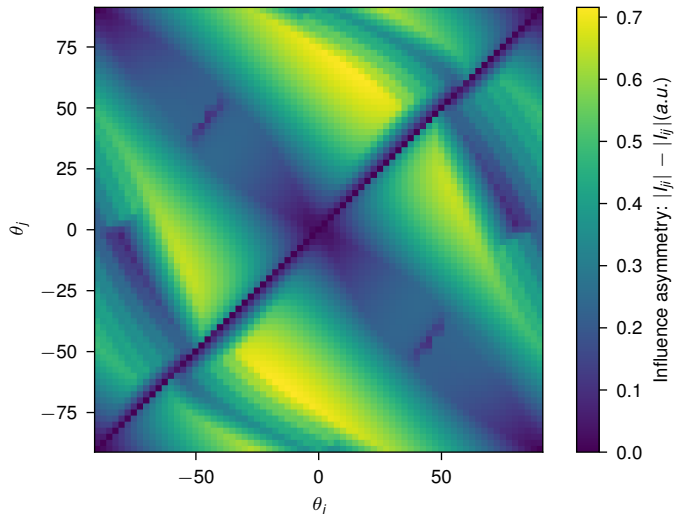

**Fig. 1**: Heatmap of the influence asymmetry $|I_{ji}| - |I_{ij}|$ between two spins $i$ and $j$, as a function of their orientation $\theta_i$ and $\theta_j$. For each $\theta_i, \theta_j$ pair, we find the external field where the influence asymmetry is maximized. In other words, the heatmap shows where the influence from $i \to j$ is greater than $j \to i$. Dark areas in the heatmap correspond to spin orientations that do not allow for asymmetry $i \to j$. We see immediately that there can be no asymmetry when the spins have the same orientation $\theta_i = \theta_j$. This is due to the mirror symmetry of the dipolar fields, where $\mathbf{h}^{ij}_{\text{dip}} = \mathbf{h}^{ji}_{\text{dip}}$. The diagonal symmetry seen in the heatmap stem from the mirror symmetry of the astroid around the horizontal axis, obtained by the transformation $(\theta_i, \theta_j) \to (-\theta_i, -\theta_j)$.

# Reversal phase diagram

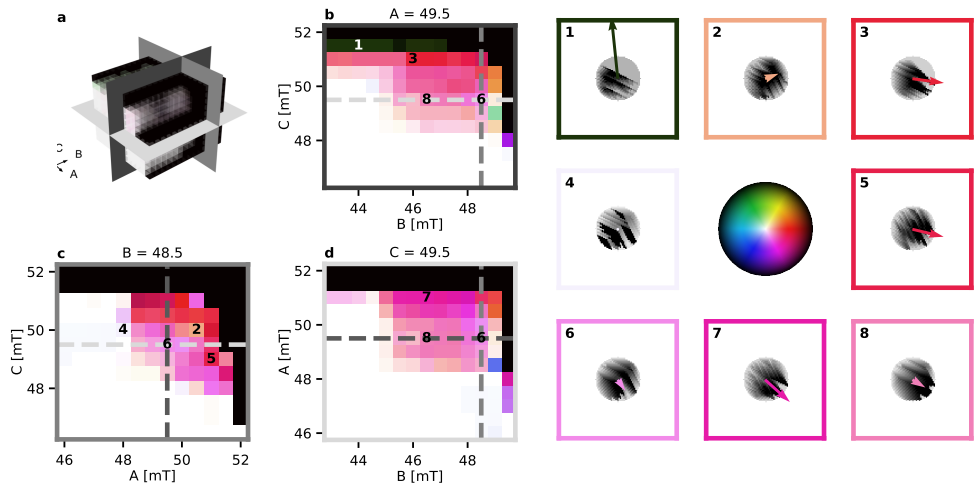

**Fig. 2**: **a** Schematic of the 3D phase space of directional reversal in geometry $T_{18}$, where hue represents direction and lightness represents the speed of growth (light is slow/no growth and dark is fast/avalanching growth). The three slices shown in **b-d** are marked in gray in **a**. 1-8 show examples of the variations of reversal we find using varying field strengths, where the reversal starts from a large circular domain and follows the gradients in the snapshots. The colored arrows show the direction of movement of the center of mass.

5

## Simulations with oppositely oriented magnetization

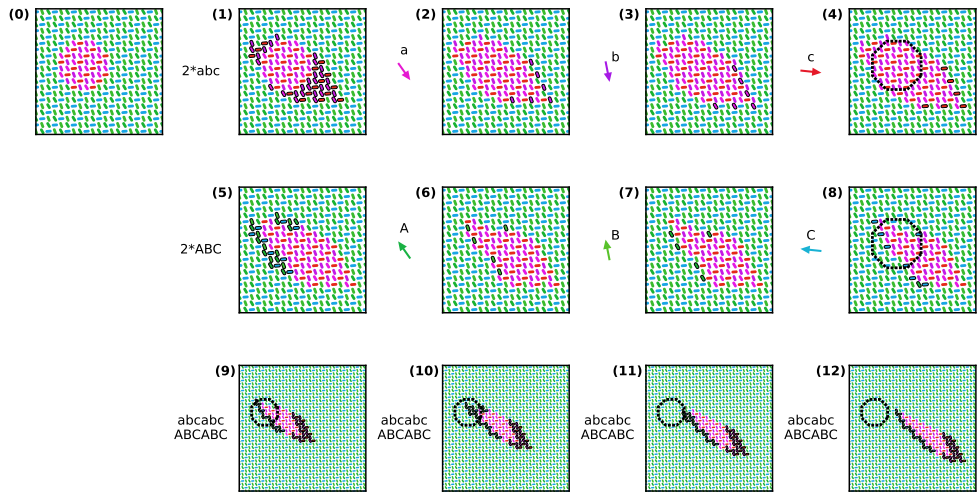

**Fig. 3**: Simulated directional domain growth with inverted magnetization. Starting from the initial state in (0) we apply three full *abc* clock cycles (1-4), growing the domain in the southeasterly direction. (1) shows the domain after the first two full *abc* cycles, while (2-4) show the domain after three subsequent field pulses, *a*, *b* and *c*. We then apply three full *ABC* cycles (5-8), reversing the domain from the northwest. By alternating cycles of growth and reversal (9-12), we can move the domain in the southeasterly direction. Note that the domain movement is in the same direction even with inverted magnetization and clock fields. The outline of the initial state is marked by the dotted line.

# XMCD-PEEM images

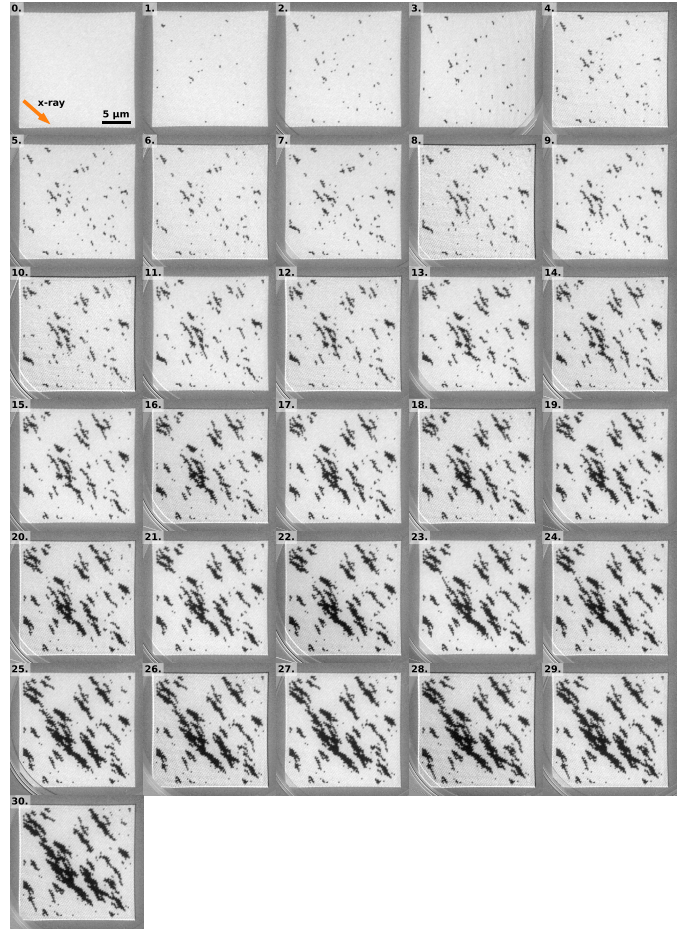

**Fig. 4**: A full series of XMCD-PEEM images showing domain growth in the southeast growth regime. Growth starts from the polarized state (white) and domains appear black. Images show the full $100 \times 100$ directional ASI based on tile $T_{18}$ consisting of $220\,\text{nm} \times 80\,\text{nm} \times 10\,\text{nm}$ stadium-shaped nanomagnets. Note that the three outermost layers of magnets are slimmer ($220\,\text{nm} \times 70\,\text{nm} \times 10\,\text{nm}$) and thus never switch. The ASI is imaged after every field pulse for 10 full $ABC$ cycles.

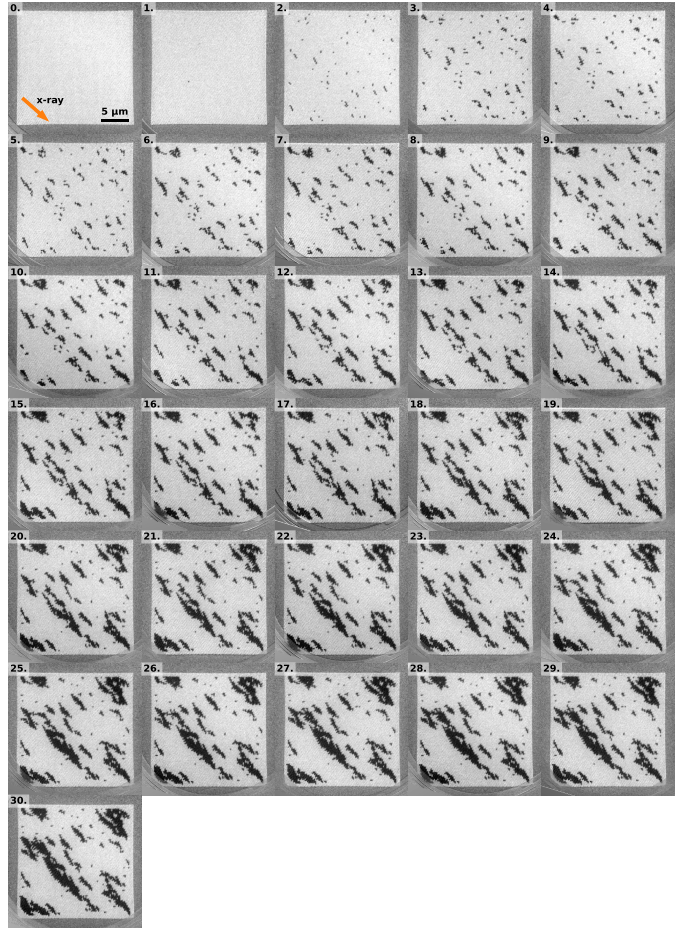

**Fig. 5**: A full series of XMCD-PEEM images showing domain growth in the northwest growth regime. Growth starts from the polarized state (white) and domains appear black. Images show the full 100 × 100 directional ASI based on tile $T_{18}$ consisting of 220 nm × 80 nm × 10 nm stadium-shaped nanomagnets. Note that the three outermost layers of magnets are slimmer (220 nm × 70 nm × 10 nm) and thus never switch. The ASI is imaged after every field pulse for 10 full $ABC$ cycles.



\end{document}